      [1999/12/01 v1.4c Il Nuovo Cimento]
\documentclass{cimento}
\usepackage{epsfig}
\usepackage{graphicx}
\def \be  {\begin{equation}}
\def \ee  {\end{equation}}  
\def \ba  {\begin{eqnarray}}
\def \ea  {\end{eqnarray}}  
\def \baa {\begin{eqnarray*}}
\def \eaa {\end{eqnarray*}}  
\def \bb  {\begin{thebibliography}}
\def \eb  {\end{thebibliography}}

\def \lab #1 {\label{#1}}

\title{Jet Production in $pp$ Collisions: Dependence on Jet
Algorithm }
\author{Asmita Mukherjee\from{ins:x}\ETC,
Werner Vogelsang\from{ins:y}}
\instlist{\inst{ins:x} Department of Physics, Indian Institute of Technology
Bombay, Powai, Mumbai 400076, India
  \inst{ins:y}Institut f\"ur Theoretische Physik,Universit\"at T\"ubingen,
72076 T\"ubingen, Germany  }

\PACSes{\PACSit{12.38.Bx}
\PACSit{13.85.Ni}}
\begin{document}

\maketitle

\begin{abstract}
We report on a recent calculation of single-inclusive high-$p_T$ jet production
in unpolarized and longitudinally polarized $pp$ collisions at RHIC, 
investigating the effect of the algorithm adopted to
define the jets on the numerical results for cross sections and spin asymmetries. 
\end{abstract}

\section{Introduction}

Jets are important tools in QCD for investigating the partonic substructure of
hadrons and interactions among partons. There is no
unique way to define a jet. 
It is thus important to compare and contrast the different available algorithms. 
The jet algorithms can be divided into two broad classes;
(i) successive combination \cite{ellis}: in this  scheme,  one defines a distance
between a pair of objects and a beam distance for every object as follows: 
\ba \label{dist1}
d_{ij} = {\mathrm{min}} ( k_{t,i}^{2 p}, k_{t,j}^{2 p}) {R_{ij}^2\over
R^2},~~~~
d_{iB}= k_{t,i}^{2 p}.     
\ea
$d_{ij}$ is called the distance between two particles $i$ and $j$ and $d_{iB}$
is the distance between the beam and the particle; $k_{t,i}$ is the
transverse
momentum of the $i$-th particle with respect to the beam direction and 
\ba\label{Rij}
R_{ij}^2= (\eta_i-\eta_j)^2 + (\phi_i-\phi_j)^2.
\ea
At each step, the smallest 
of all distances is determined. If it is a beam distance, the object is 
called a jet and is removed from the event; otherwise the two objects $j,k$
are combined into a single one. Examples of successive combination algorithms
are the $k_t$ algorithm \cite{catani1}, where  $p=1$,  and the anti-$k_t$ 
algorithm \cite{css} for which $p=-1$.
   
(ii) Cone algorithms: in these algorithms the jet is defined in terms of stable  
cones as circles of fixed radius in the $\eta$-$\phi$ plane, 
such that the sum of the $4$-momenta of the particles in it points
in the direction of the center of the cone. One defines the jet by all
 particles $j$ that satisfy \cite{cone} 
\begin{equation}
\label{eq:conedef}
R_{jJ}^2\equiv (\eta_J-\eta_j)^2 + (\phi_J-\phi_j)^2 \le R^2 ,
\end{equation}
where $\eta_J$ and $\phi_J$ are the pseudo-rapidity and azimuthal angle of
the jet, respectively.
Higher order QCD corrections are important  
as the dependence on the factorization and
renormalization  
scales is expected to be reduced when the corrections are
included. 
In the case of jet production, higher order corrections are particularly
important, as only
at NLO the QCD structure of the jet starts to play a role in the
theoretical     
description of the process. In fact, some of the popular cone algorithms    
are known to be collinear and infra-red unsafe at NNLO or when
multiple jets
are considered. 

Single inclusive large $p_T$ jets in longitudinally polarized  
$pp$ collisions at RHIC are important tools to gain access to
the polarized gluon distribution in the nucleon. 
The cross section for single inclusive jet production at 
RHIC has been calculated at NLO using a Monte-Carlo technique \cite{jet1} 
in the cone algorithm.
However a largely analytic technique was developed in \cite{jet2} in the 
limit when the cone
opening is   
relatively small (small cone approximation). This is advantageous because it
leads      
to much faster and more efficient computer codes as the singularities in the        
intermediate steps cancel analytically and one does not have to treat them through
delicate      
numerical techniques. The basis of such an analytic calculation is the
observation      
that the inclusive jet production proceeds through the same partonic
subprocesses   
as single inclusive hadron production and it is possible to convert an NLO
cross section
for single inclusive hadron production to the one for jet
production.     
The main difference between  the two cases is the fact that in single
inclusive     
hadron production, one integrates over the full phase space of the
unobserved       
partons. This leads to collinear singularities, which are absorbed in the
parton to hadron fragmentation functions. In contrast, for a jet,
final-state particles that move in roughly the same direction 
will jointly produce the jet. This makes the cross section more
inclusive, and (for a proper jet definition) final state
singularities must cancel. The cross section for single
inclusive  
hadron production can however be transformed into that for single inclusive jet
production \cite{jet2}. In the limit of small cone size, this 
transformation can even be performed analytically. 

We have recently extended the above analytic technique to the more
widely used successive combination schemes (for example, 
$k_t$ or anti-$k_t$), assuming
again that
the jet parameter $R$ used to define the distance between two objects in
this algorithm is not too large \cite{jetalgo}. When systematically expanded 
around $R=0$, the dependence of the partonic cross
sections on $R$ is of the form ${\cal A}
\log R+{\cal B}+{\cal O} (R^2)$.   
The coefficients ${\cal A}$ and ${\cal B}$ are  calculated
analytically, and the remaining terms ${\cal O} (R^2)$ and beyond are neglected. 
We refer to this approximation as ``Narrow Jet
Approximation'' (NJA). The NJA gives a very accurate description of 
the single inclusive jet cross sections at RHIC, Tevatron and even at the LHC \cite{jetalgo}. It turned out that the cross 
sections for single inclusive jet production in the cone and
the successive recombination algorithms differ by 
calculable finite terms. We have also given numerical estimates of the cross section both for
unpolarized and 
longitudinally polarized collisions at RHIC, and examined the
effect of the choice of jet algorithm 
on the double longitudinal spin asymmetry. Here we give a brief report 
of \cite{jetalgo}.

\section{Cross section for single inclusive jet production in $pp$ collisions} 

We consider single-inclusive jet production in hadronic collisions, 
$pp \rightarrow {\mathrm{jet}}\,X$, where the jet has a transverse momentum
$p_{T_J}$,
rapidity $\eta_J$, and azimuthal angle $\phi_J$. Note that on top of the
choice of jet algorithm one also has to define how objects
are to be merged to form the jet. We choose
to define the four-momentum of the jet as the sum 
of four-momenta of the partons that form the jet for both algorithms (``$E$ 
recombination scheme''~\cite{cone}).

In order to calculate the single-inclusive jet
cross section at NLO, we start from the NLO 
single-parton inclusive cross sections $d\hat\sigma_{ab\rightarrow c X}$,
relevant for single-inclusive hadron production process $pp\to hX$ and 
analytically known.  
For a jet cross section, the observed final state should 
not  be given by parton $c$ only, but by partons $c$ and $d$ jointly, when
the two are close to each other (as two
partons together can form the jet). In order to calculate this 
one first considers a ``jet cone'' characterized by a jet parameter $R$ 
around the observed parton $c$ and notices that in the NLO single-parton 
inclusive cross section there is a configuration where 
an additional parton $d$ is inside the cone 
(we use the term ``cone'' for simplicity, the considerations apply to any 
jet definition). One subtracts these 
contributions and replaces them by terms for which partons $c$ and $d$ are
both inside the cone and  form the observed jet together. For a given partonic
process $ab\to cde$ we then have,
\begin{eqnarray}
d \hat{\sigma}_{ab\rightarrow \mathrm{jet}X} &=&
[d \hat{\sigma}_c -d \hat{\sigma}_{c(d)}-
d \hat{\sigma}_{c(e)}]\nonumber\\[1mm] &+&
 [d \hat{\sigma}_d -d \hat{\sigma}_{d(c)}-
d \hat{\sigma}_{d(e)}]\nonumber\\[1mm] &+&
[d  \hat{\sigma}_e -d \hat{\sigma}_{e(c)}-
d \hat{\sigma}_{e(d)}]\nonumber\\[1mm] &+&
d \hat{\sigma}_{cd} + d \hat{\sigma}_{ce}+
d \hat{\sigma}_{de} .
\label{jetform}
\end{eqnarray} 
Here $d \hat{\sigma}_j$ is the single-parton inclusive cross section where
parton $j$
is observed (which also includes the virtual corrections),  $d
\hat{\sigma}_{j(k)}$  is the cross section 
where parton $j$ is observed but parton $k$ is also in the cone, and
$d \hat{\sigma}_{jk}$ is the cross 
section when both partons $j$ and $k$ are inside the cone and jointly form
the jet. One has to note that the single-parton inclusive cross section
contains a  subtraction of final-state collinear singularities in the
modified minimal subtraction ($\overline{\mathrm{MS}}$) scheme.
One has to perform an $\overline{\mathrm{MS}}$ subtraction also of the singularities
in the  $d \hat{\sigma}_{j(k)}+d \hat{\sigma}_{k(j)}-d \hat{\sigma}_{jk}$.

The difference between the cone and $k_t$ type algorithms resides
entirely in the 
 $d \hat{\sigma}_{jk}$, which can be calculated analytically in the NJA. 
The reason for this difference is as follows: for
the $k_t$-type algorithms the two partons $j,k$ are merged into one jet if
their distance
defined in~(\ref{dist1}) is smaller than their respective beam distances
$d_{iB}$ and $d_{jB}$
defined in~(\ref{dist1}). For $d \hat{\sigma}_{jk}$ this has to hold, 
and we arrive at the condition
\be\label{j1}
R_{jk}^2\leq R^2 \quad \quad {\mathrm{for}}\; k_t{\mathrm{-type
\;algorithms}},
\ee
with $R_{jk}$ defined in Eq.~(\ref{Rij}). This condition
is true  for {\it all}
$k_t$-type algorithms. Whereas in cone algorithm,  
Eq.~(\ref{eq:conedef}) is valid:
\be\label{j2}
R_{jJ}^2 \leq R^2\;  \wedge\; R_{kJ}^2 \leq R^2 \;\;
\quad {\mathrm{for\; cone \;algorithm}}.
\ee
We find that the difference in the cross sections calculated for the two
algorithms is finite, as it must be.

\section{Numerical Results}

\begin{figure}[t]
\vspace*{3mm}
\centering   
\includegraphics[width=6cm,clip]{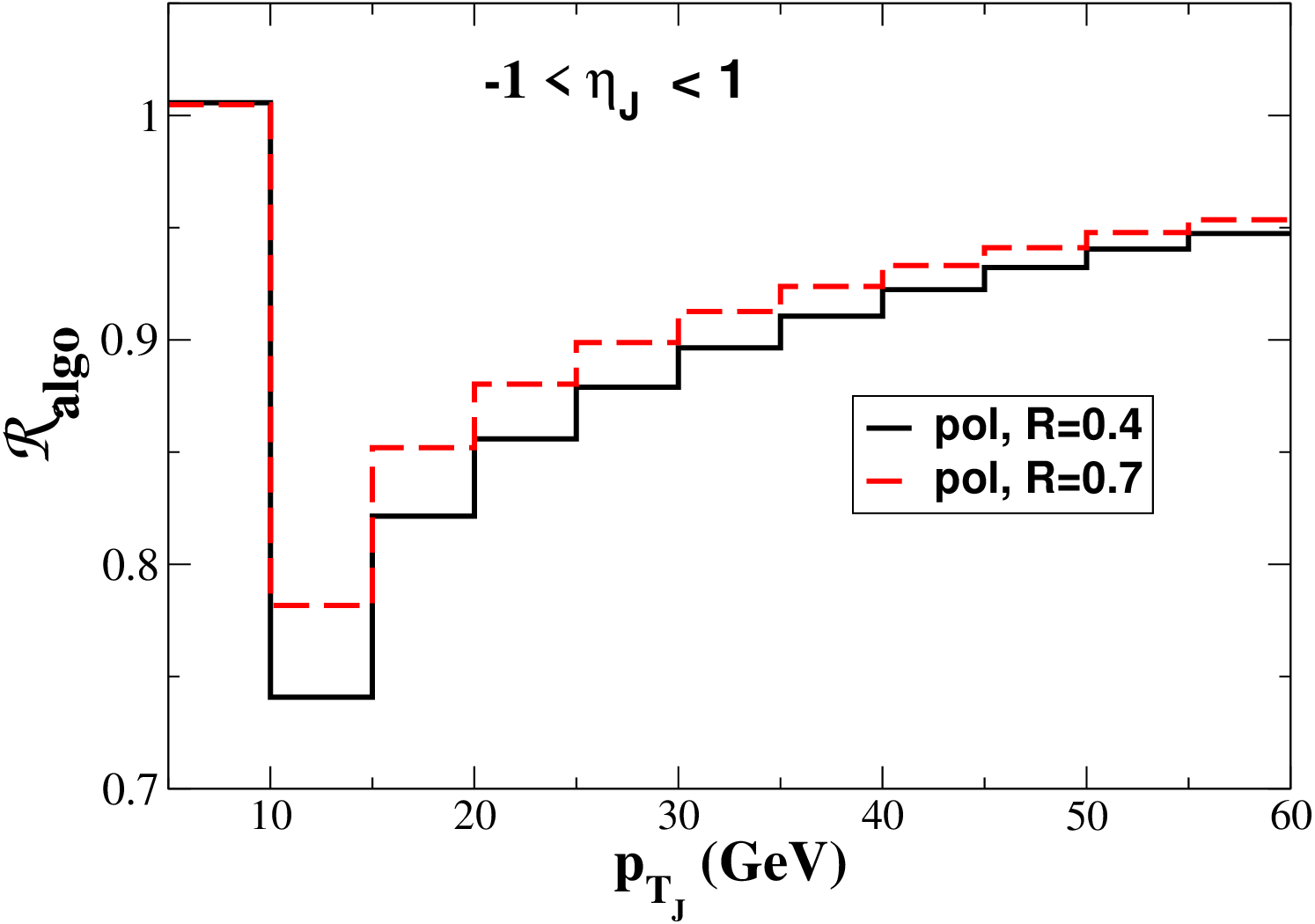}
\hspace{0.2cm}
\includegraphics[width=6cm,clip]{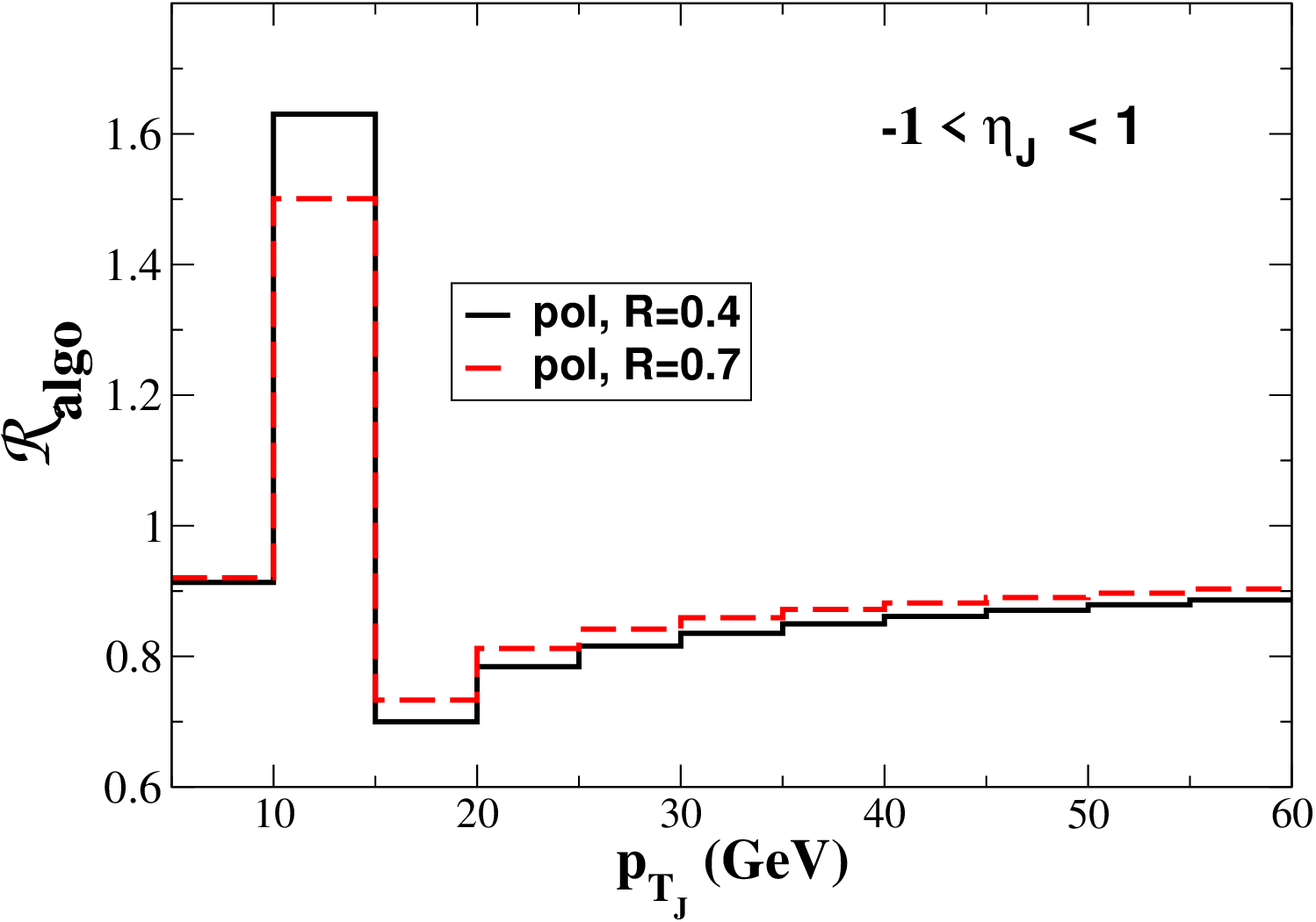}
\caption{\label{fig:ratio2pol}
\sf The ratio ${\cal R}_{\mathrm{algo}}$ at RHIC for
$\sqrt{S} = 200$ GeV (left) and $\sqrt{S} = 
500$ GeV (right), for the spin-dependent case.
Results are shown for two different values of the jet parameter $R$. We   
have chosen the factorization and renormalization scales as $\mu_F=\mu_R=p_{T_J}$.}

\end{figure}  

Next, we present some numerical results for single-inclusive jet
production cross sections and spin asymmetries in $pp$ collisions at RHIC. 
We use the CTEQ6.6M parton distributions~\cite{cteq66} for the
unpolarized cross section and the ``DSSV'' helicity parton distributions
of Ref.~\cite{dssv} for the polarized case. 
We define the ratio
\be
{\cal R}_{\mathrm{algo}} \equiv 
\frac{\left[d^2(\Delta)\sigma/d p_{T_J}d\eta_J\right]_{k_t{\mathrm{-type}}}}
{\left[d^2(\Delta)\sigma/d p_{T_J}d\eta_J\right]_{{\mathrm{cone}}}},
\ee
where the jet parameter $R$ is the same for  both cross sections.

Figure~\ref{fig:ratio2pol} shows
the ratio ${\cal R}_{\mathrm{algo}}$ for polarized collisions at RHIC,
calculated for the factorization and renormalization 
scales as $\mu_F=\mu_R= p_{T_J}$, as a function of $p_{T_J}$ in bins of $p_{T_J}$ , for $R=0.4$ and $0.7$.
We present results for two values of c. m. s. energies at RHIC, 
$\sqrt{S}=200$~GeV (left) and $\sqrt{S}=500$~GeV (right) . ${\cal R}_{\mathrm{algo}}$
is around 90\% at high  $p_{T_J}$, 
but deviates largely from one in the bin around  $p_{T_J}=12.5$~GeV.
The reason for this is that for the DSSV set of parton distributions the
polarized jet cross section changes sign around $p_{T_J}=10$~GeV.
Depending on the jet algorithm, the zero is  at a slightly different value of
$p_T$. This shows that in regions where the polarized cross
section is very small it is very sensitive  to the choice of jet algorithm.

\begin{figure}[t!]
\vspace{0.6cm}
\centering
\includegraphics[width=5cm]{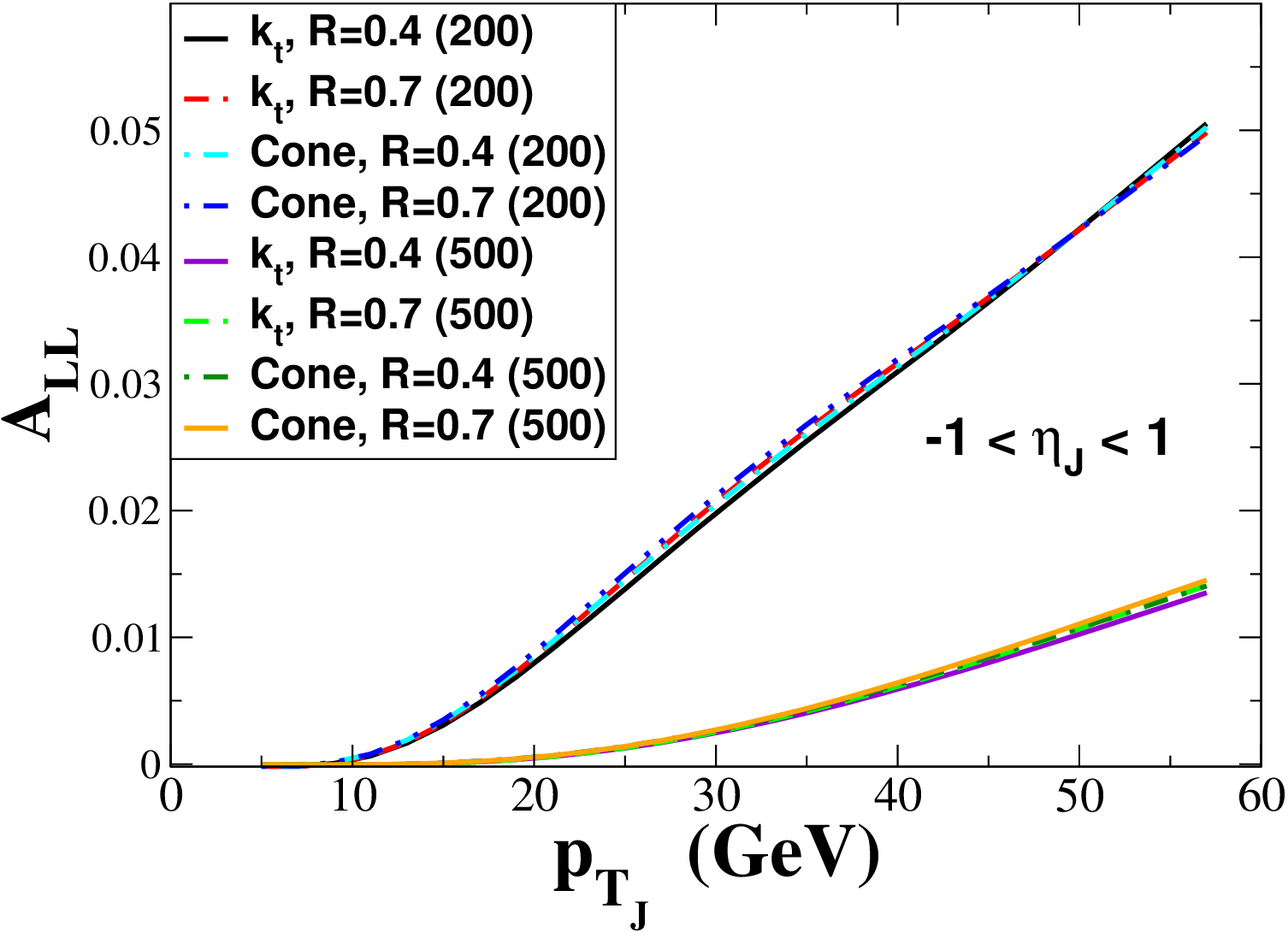}
\caption{\label{fig:asym}
\sf Double-longitudinal spin asymmetries $A_{LL}$ at RHIC, for
$\sqrt{S}=200$~GeV    
and $\sqrt{S}=500$~GeV and various jet definitions. We have averaged over 
$|\eta_J|\leq 1$.
The factorization and renormalization scales have been chosen to be $\mu_F=\mu_R=p_{T_J}$. }
\end{figure}  

The double longitudinal spin asymmetries $A_{LL}$ at RHIC are
defined by
\be\label{alldef}
A_{LL}\equiv\frac{d^2\Delta\sigma/dp_{T_J}d\eta_J}{d^2\sigma/dp_{T_J}d\eta_J}.
\ee 
For the denominator we use the spin-averaged cross sections and for the
numerator the polarized ones. The results are shown in Fig.~\ref{fig:asym}.
As one can see, the asymmetries 
are quite insensitive to the jet algorithm chosen, and also to the value of
the jet parameter $R$ for all values of $p_{T_J}$ where the asymmetry is
sizable. Our results are useful for the analysis of the data on the double longitudinal
spin asymmetry in single inclusive jet production by the STAR collaboration at
RHIC \cite{star}.

\acknowledgments

AM thanks the Alexander von Humboldt Foundation, Germany, for support
through a Fellowship for Experienced Researchers. 
AM thanks the organizers of QCD-N12, Bilbao, Spain for the
kind invitation.


\begin{thebibliography}{0}

\bibitem{ellis} \BY{S.~D.~Ellis and D.~E.~Soper}
\IN{Phys.\ Rev.\ D}{48}{1993}{3160}; [hep-ph/9305266].

\bibitem{catani1}
\BY{S.~Catani, Y.~L.~Dokshitzer, M.~H.~Seymour and B.~R.~Webber}
\IN{Nucl.\ Phys.\ B}{406}{1993}{187}, {\it and references therein}.

\bibitem{css} \BY{M.~Cacciari, G.~P.~Salam and G.~Soyez}
  \IN{JHEP}{0804}{2008}{063};  [arXiv:0802.1189 [hep-ph]].

\bibitem{cone}  \BY{G.~C.~Blazey {\it et al.}}
  hep-ex/0005012, {\it and references therein}.

\bibitem{jet1} \BY{D.~de Florian, S.~Frixione, A.~Signer and W.~Vogelsang}
\IN{Nucl.\ Phys.\ B}{539}{1999}{455};
[hep-ph/9808262].
  
\bibitem{jet2} \BY{ B.~J\"{a}ger, M.~Stratmann and W.~Vogelsang}
 \IN{Phys.\ Rev.\ D}{70}{2004}{034010};
  [hep-ph/0404057].

\bibitem{jetalgo} \BY{A. Mukherjee, W. Vogelsang}
  \IN{Phys. Rev. D}{86}{2012}{094009}; [arXiv:1209.1785 [hep-ph]].


\bibitem{cteq66}
\BY{P.~M.~Nadolsky {\it et al.}}
  \IN{Phys.\ Rev.\ D}{78}{2008}{013004};
  [arXiv:0802.0007 [hep-ph]].

\bibitem{dssv} \BY{D.~de Florian, R.~Sassot, M.~Stratmann and W.~Vogelsang}
  \IN{Phys.\ Rev.\ Lett.}{101}{2008}{072001}; [arXiv:0804.0422 [hep-ph]]; 
\IN{Phys.\ Rev.\ D}{80}{2009}{034030};
  [arXiv:0904.3821 [hep-ph]].

\bibitem{star}
\BY{J.~Kapitan} [STAR Collaboration],
  arXiv:1111.1892 [nucl-ex];
\BY{E.~Bruna} [STAR Collaboration],
  \IN{AIP Conf.\ Proc.}{1422}{2012}{190}.
     


\end{thebibliography}
\end{document}